\documentclass[fleqn]{2020SCGE}
\begin{document}

\ensubject{subject}

\ArticleType{Article}
\Year{2020}
\Month{August}
\Vol{63}
\No{8}
\DOI{https://doi.org/10.1007/s11433-019-1516-y}
\ArtNo{280411}
\ReceiveDate{November 11, 2019}
\AcceptDate{January 21, 2020}
\OnlineDate{April 13, 2020}

\title{Forecast for weighing neutrinos in cosmology with SKA}{Forecast for weighing neutrinos in cosmology with SKA}

\author[1]{Jing-Fei Zhang}{}%
\author[1]{Bo Wang}{}
\author[1,2,3]{Xin Zhang}{{zhangxin@mail.neu.edu.cn}}


\AuthorMark{J.-F. Zhang}

\AuthorCitation{J.-F. Zhang, B. Wang, X. Zhang}

\address[1]{Department of Physics, College of Sciences, Northeastern
University, Shenyang 110819, China;}
\address[2]{Ministry of Education's Key Laboratory of Data Analytics and Optimization
for Smart Industry, Northeastern University, Shenyang 110819, China}
\address[3]{Center for High Energy Physics, Peking University, Beijing 100080, China}


\abstract{We investigate what role the SKA neutral hydrogen (HI) intensity mapping (IM) and galaxy sky surveys will play in weighing neutrinos in cosmology. We use the simulated data of the baryon acoustic oscillation (BAO) measurements from the HI surveys based on SKA1 (IM) and SKA2 (galaxy) to do the analysis. For the current observations, we use the {\it Planck} 2015 cosmic microwave background (CMB) anisotropies observation, the optical BAO measurements, the type Ia supernovae (SN) observation (Pantheon compilation), and the latest $H_0$ measurement. We consider three mass ordering cases for massive neutrinos, i.e., the normal hierarchy (NH), inverted hierarchy (IH), and degenerate hierarchy (DH) cases. It is found that the SKA observation can significantly improve the constraints on $\Omega_{\rm m}$ and $H_0$. Compared to the current observation, the SKA1 data can improve the constraints on $\Omega_{\rm m}$ by about 33\%, and on $H_0$ by about 36\%; the SKA2 data can improve the constraints on $\Omega_{\rm m}$ by about 58\%, and on $H_0$ by about 66\%. It is also found that the SKA observation can only slightly improve the constraints on $\sum m_\nu$. Compared to the current observation, the SKA1 data can improve the constraints on $\sum m_\nu$ by about 4\%, 3\%, and 10\%, for the NH, IH, and DH cases, respectively; the SKA2 data can improve the constraints on $\sum m_\nu$ by about 7\%, 7\%, and 16\%, for the NH, IH, and DH cases, respectively.}

\keywords{21 cm observation, intensity mapping survey, Square Kilometre Array, baryon acoustic oscillations, neutrino mass}

\PACS{95.36.+x, 98.80.Es, 98.80.-k}

\maketitle


\begin{multicols}{2}
\section{Introduction}\label{section1}

Precise measurement of cosmological parameters is one of the core tasks of cosmology research. The answers to almost all important scientific questions in cosmology depend on precise measurements of cosmological parameters. The measurement of cosmological parameters is closely related to the cosmological model because the determination of cosmological parameters is usually done by fitting observational data under the assumption of a specific cosmological model.

After decades of development, the study of cosmology has
\Authorfootnote
entered the era of precision cosmology. At present, a standard model of cosmology has been basically established, of which the prototype is the so-called $\Lambda$ cold dark matter ($\Lambda$CDM) model. The basic version of the $\Lambda$CDM model has only 6 base parameters, and this base $\Lambda$CDM model is favored by the {\it Planck} observation of cosmic microwave background (CMB) anisotropies \cite{Aghanim:2018eyx}. However, the base $\Lambda$CDM cosmology has also encountered some serious challenges in the aspect of observation. Some significant tensions occur between different observations based on the base $\Lambda$CDM cosmology, such as the Hubble constant tension and the matter density fluctuation amplitude tension \cite{Ade:2013zuv,Ade:2015xua}. This actually indicates that the standard $\Lambda$CDM model needs to be extended. Of course, extra parameters need to be introduced in extended cosmological models.

The current mainstream cosmological probes mainly include the following ones: the CMB anisotropy (temperature and polarization) power spectra measurement, the baryon acoustic oscillation (BAO) measurement, the type Ia supernova (SN) observation, the direct measurement of the Hubble constant $H_0$, the shear measurement of weak gravitational lensing, the galaxy cluster counts, the redshift space distortion (RSD) measurement, and the lensing measurement of CMB. These cosmological probes make precise measurements of the expansion history of the universe (the first four) and the structural growth of the universe (the last four). These probes are all based on the optical measurements, and will be further greatly developed in the future. However, in the future we also need to develop new cosmological probes other than those sky surveys based on the optical measurements. Actually, the gravitational-wave standard siren observation and the radio 21 cm observation are thought to be the most important new cosmological probes in the forthcoming years \cite{Zhang:2019ylr}. Undoubtedly, the Square Kilometre Array (SKA) radio telescope \footnote{https://www.skatelescope.org} will definitely play a crucial role in the studies of astronomy and cosmology.

SKA is the largest synthetic aperture radio telescope all over the world that the international astronomy community plans to build, with the effective receiving area of one square kilometer \cite{Wu:2019}. SKA devotes itself to answering some of the most fundamental questions about the universe. Research on neutral hydrogen (HI) 21 cm cosmology is one of the important breakthroughs that we are committed to making. The SKA HI sky survey based on the 21 cm intensity mapping (IM) technique can be used to measure HI power spectrum, and the related BAO and RSD signals, of which the accuracies can reach or exceed those of future large optical sky survey projects, which will play an important role in the measurement of cosmological parameters and the exploration of the nature of dark energy. For constraints on the equation-of-state (EoS) parameter of dark energy using the simulated data based on SKA observation, see refs.~\cite{Bull:2015nra,Raccanelli:2015qqa,Zhao:2015wqa,Bacon:2018dui,Chen:2019jms,Liu:2019asq,Zhang:2019dyq,Yohana:2019ahg}. (For investigations on precise measurements of inflationary features with 21 cm simulated observations from Tianlai and SKA, see refs.~\cite{Xu:2014bya,Xu:2016kwz}.) As a next step, we wish to see if the SKA observation could play a significant role in helping improve the constraints on the total neutrino mass in cosmology.

Neutrino oscillation experiments have revealed the fact that neutrinos have masses, but the neutrino oscillation experiments cannot measure the absolute masses of neutrinos. Neutrino masses play an important role in the cosmic evolution, which not only affect the expansion history of the universe, but also affect the structural growth of the universe. The {\it Planck} observations, combined with other astrophysical observations, have so far constrained the total mass of neutrinos to be less than about 0.12 eV \cite{Aghanim:2018eyx}. It should be pointed out that recent research has shown that the 10-year observation of gravitational-wave standard sirens from the Einstein Telescope (ET) can help improve the constraints on the total neutrino mass by about 10\% \cite{Wang:2018lun}. In this work, we will use the simulated data of the HI sky survey observations of SKA to perform constraints on the total neutrino mass, of which the purpose is to see whether the SKA observations could help improve the cosmological measurement of the neutrino mass.

This paper is organized as follows. In sect.~\ref{sec:2}, we introduce the analysis method and the observational data used in this work. We will consider the actual observational data of the current mainstream cosmological probes, and use these data to constrain the cosmological model with massive neutrinos. Then, we will further consider the simulated data of the HI sky survey observations of SKA in the cosmological fit. The neutrino mass splittings measured in the neutrino oscillation experiments are considered in this work. In sect.~\ref{sec:3}, we report the constraint results and make some relevant discussions. The conclusion of this work is given in sect.~\ref{sec:4}.

\section{Method and data}\label{sec:2}
In this work, we consider massive neutrinos in a flat $\Lambda$CDM model. With the increasing accuracy of the observational data, the effects from the mass splittings of the neutrinos can gradually be sensitive to the observational data, and thus we also consider the mass splittings of the neutrinos in this work. In the neutrino oscillation experiments, the solar and reactor experiments have measured $\Delta m_{21}^2\simeq 7.5\times 10^{-5}$ eV$^2$, and the atmospheric and accelerator beam experiments have measured $|\Delta m_{31}^2|\simeq 2.5\times 10^{-3}$ eV$^2$ \cite{pdg}, indicating that there are two possible mass orders, i.e., the normal hierarchy (NH) with $m_1<m_2\ll m_3$ and the inverted hierarchy (IH) with $m_3\ll m_1<m_2$. Thus, in this paper, we follow refs.~\cite{Huang:2015wrx,Wang:2016tsz,Xu:2016ddc,Feng:2017mfs,Zhao:2017jma ,Guo:2018gyo,Guo:2018ans,Feng:2019mym,Vagnozzi:2017ovm,Feng:2019jqa} to use these two mass splittings as an input to parameterize the total mass of neutrinos, in terms of $m_1$ as a free parameter for the NH case and in terms of $m_3$ as a free parameter for the IH case. In addition, we also consider the degenerate model of the neutrino mass with $m_1=m_2=m_3$ in this work, and this case is called degenerate hierarchy (DH) following the literature. Note that there are lower bounds of the total mass for the NH case and the IH case, which are 0.06 and 0.10 eV, respectively. For the DH case, there is no a lower bound of the total mass.

We will first use the current observations to constrain the cosmological model involving massive neutrinos. Then, we will use the simulated data of the HI sky survey observations of SKA to constrain the same model, and we will directly see how the SKA observations would help improve the constraints on the neutrino mass.

The current observations we use in this paper are the CMB, BAO, SN, and $H_0$ data. For the CMB data, we use the {\it Planck} temperature and polarization power spectra at the full range of multipoles \cite{Aghanim:2015xee}, which is denoted as ``{\it Planck} TT,TE,EE+lowTEB''. For the optical BAO data, we use the measurements from the six-degree-field galaxy survey (6dFGS) at $z_{\rm eff}=0.106$ \cite{Beutler:2011hx}, the SDSS main galaxy sample (MGS) at $z_{\rm eff}=0.15$ \cite{Ross:2014qpa}, the SDSS DR12 galaxy sample at $z_{\rm eff}=0.38$, $z_{\rm eff}=0.51$, and $z_{\rm eff}=0.61$ \cite{Alam:2016hwk}. For the SN data, we use the latest sample consisting of 1048 data from the Pantheon compilation \cite{Scolnic:2017caz}. For the $H_0$ data, we use the latest measurement given by ref.~\cite{Riess:2018byc}, with the result  $H_0=73.52\pm1.62\ {\rm km\ s^{-1}\ Mpc^{-1}}$. The basic data combination used in this work is from the current observation, i.e., CMB+BAO+SN+$H_0$, denoted as ''data0'' for convenience in the following discussions.

We consider the forecasted BAO data of the neutral hydrogen survey observations from the SKA Phase 1 (denoted as SKA1) and the SKA Phase 2 (denoted as SKA2). SKA1 is comprised of two telescopes, i.e., SKA1-MID and SKA1-LOW. SKA1-MID is a mid-frequency dish array located in South Africa, observing the radio frequencies of 0.35 -- 1.75 GHz, and SKA1-LOW is located in the western of Australia, observing the radio frequencies of 0.05 -- 0.35 GHz \cite{Bacon:2018dui}. SKA1-MID will operate in two frequency bands, with Band1 observing at 350 -- 1050 MHz and Band2 observing at 950 -- 1750 MHz. SKA2 will perform an immense galaxy redshift survey over three quarters of the sky, with an impressive sensitivity surpassing almost all the other planned BAO measurements at the redshift range of $0.4 \lesssim z \lesssim 1.3$.  We will mainly consider SKA1-MID and SKA2 in this work, since SKA1-MID contains the main frequency range for IM and they are both able to observe the large-scale structure at low redshifts $0\lesssim z \lesssim 3$ where dark energy dominates the evolution of the universe.

The expected BAO measurements by the SKA have been forecasted in ref.~\cite{Bull:2015nra} based on the Fisher forecasting formalism developed in ref.~\cite{Bull:2014rha}. For SKA1-MID (IM), the experimental specifications used in the forecast are given in Table 2 of ref.~\cite{Bull:2015nra}. For SKA2, only a galaxy survey is considered in the forecast, although actually an IM survey based on a mid-frequency aperture array should be able to provide similar BAO signals out to $z\simeq 2$; the number counts used in the forecast can be found in Table 1 of ref.~\cite{Bull:2015nra}. The expected relative errors of $H(z)$ and $D_A(z)$ in the BAO measurements by the SKA, using the Fisher forecasting method, are given in Figure 3 of ref.~\cite{Bull:2015nra}.

In this work, we directly use this forecasted HI BAO data of SKA presented in ref.~\cite{Bull:2015nra} to make an analysis for the cosmological parameter constraints. In the cosmological fit, we use the Markov-chain Monte Carlo (MCMC) method, rather than the Fisher matrix method, to infer the posterior probability distributions of cosmological parameters. We use these forecasted $H(z)$ and $D_A(z)$ data to establish likelihood functions for SKA1 and SKA2. Here we note that, in the cosmological fit: (i) we only consider the BAO measurements, but not consider the RSD measurements; (ii) for the BAO measurements, the assumption of no correlation between $H(z)$ and $D_A(z)$ is made. We thus consider another two data combinations in this work: CMB+BAO+SN+$H_0$+SKA1, denoted as ''data1'', and CMB+BAO+SN+$H_0$+SKA2, denoted as ''data2''.

In the late universe, the Hubble expansion rate $H(z)$ is given by the Friedmann equation $H(z)=H_0[(1-\Omega_{\rm m})+\Omega_{\rm m}(1+z)^3]^{1/2}$. The energy density of radiation can be neglected in the late times. The energy density of matter is contributed from baryons, cold dark matter, and massive neutrinos, i.e., $\Omega_{\rm m}=\Omega_{\rm b}+\Omega_{\rm c}+\Omega_\nu$. The fractional density of massive neutrinos $\Omega_\nu$ is related to their total mass via the relationship $\Omega_\nu=\sum m_\nu/(94.1h^2~{\rm eV})$, where $\sum m_\nu$ is the total neutrino mass and $h$ is the dimensionless Hubble constant ($H_0=100h\ {\rm km\ s^{-1}\ Mpc^{-1}}$). The angular diameter distance $D_A(z)$ can be calculated through the formula $D_A(z)=(1+z)^{-1}\int_0^z dz'/H(z')$ from a specific cosmological model.

We use the code package {\tt CosmoMC} \cite{Lewis:2002ah} based on the MCMC method to infer the posterior probability distributions of parameters and their best-fit values and errors. For each neutrino mass ordering case of the cosmological model, we use the three data combinations to perform constraints and to estimate parameters.
\begin{table*}[!htp]
\footnotesize
\caption{\label{tab1} The constraint results for the NH case of the neutrino mass ordering. Here, data0 denotes CMB+BAO+SN+$H_0$, data1 denotes CMB+BAO+SN+$H_0$+SKA1, and data2 denotes CMB+BAO+SN+$H_0$+SKA2. Note that $\sum m_\nu$ is in units of ${\rm eV}$ and $H_0$ is in units of ${\rm km\ s^{-1}\ Mpc^{-1}}$.}
\tabcolsep 35.6pt%
\begin{tabular}{c c c c}
\toprule
Data&data0&data1&data2\\
\hline
$\Omega_bh^2$&$0.02241\pm0.00014$&$0.02241\pm0.00012$&$0.02241\pm0.00011$\\
$\Omega_ch^2$&$0.1177\pm0.0010$&$0.11772\pm0.00082$&$0.11773^{+0.00070}_{-0.00058}$\\
$100\theta_{\rm{MC}}$&$1.04102\pm0.00030$&$1.04101\pm0.00028$&$1.04101\pm0.00026$\\
$\tau$&$0.091\pm0.017$&$0.091\pm0.016$&$0.091\pm0.016$\\
${\rm{ln}}(10^{10}A_s)$&$3.113\pm0.033$&$3.113\pm0.032$&$3.112\pm0.032$\\
$n_s$&$0.9700\pm0.0040$&$0.9700\pm0.0036$&$0.9699\pm0.0035$\\
\hline
$\Omega_m$&$0.3054\pm0.0061$&$0.3052\pm0.0041$&$0.3052\pm0.0025$\\
$H_0$&$67.99\pm0.47$&$68.00\pm0.30$&$68.00\pm0.16$\\
\hline
$\sum m_\nu$&$<0.148$&$<0.142$&$<0.137$\\
\bottomrule
\end{tabular}
\centering
\end{table*}

\begin{figure*}[!t]
\centering
\includegraphics[scale=0.50]{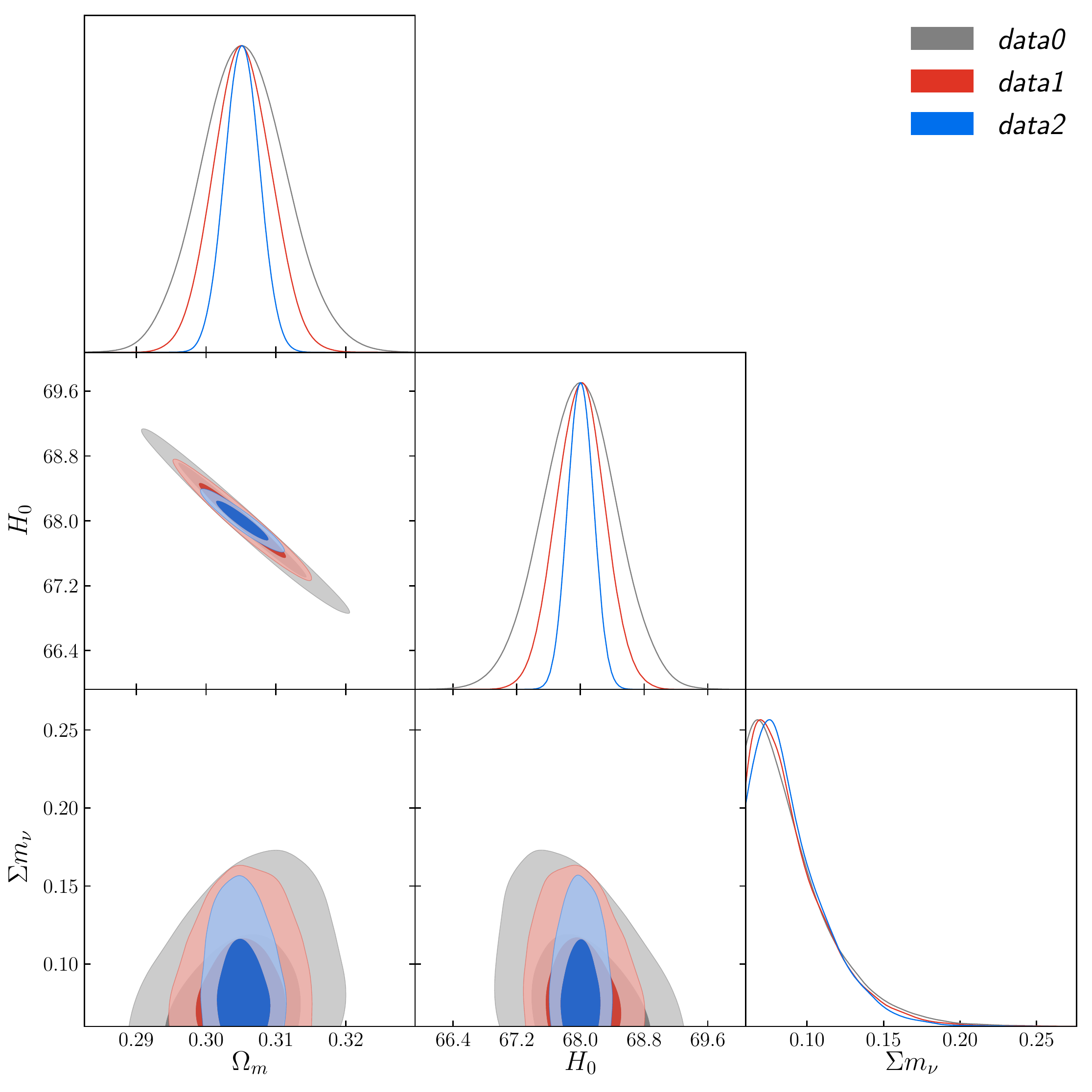}
\caption{\label{fig1} The constraint results for the NH case of the neutrino mass ordering. Marginalized one- and two-dimensional posterior distributions of $\sum m_\nu$, $\Omega_{\rm m}$, and $H_0$ using data0, data1, and data2. Here, data0 denotes CMB+BAO+SN+$H_0$, data1 denotes CMB+BAO+SN+$H_0$+SKA1, and data2 denotes CMB+BAO+SN+$H_0$+SKA2. Note that $\sum m_\nu$ is in units of ${\rm eV}$ and $H_0$ is in units of ${\rm km\ s^{-1}\ Mpc^{-1}}$.}
\end{figure*}

\begin{table*}[!t]
\footnotesize
\centering
\caption{\label{tab2}The constraint results for the IH case of the neutrino mass ordering. Here, data0 denotes CMB+BAO+SN+$H_0$, data1 denotes CMB+BAO+SN+$H_0$+SKA1, and data2 denotes CMB+BAO+SN+$H_0$+SKA2. Note that $\sum m_\nu$ is in units of ${\rm eV}$ and $H_0$ is in units of ${\rm km\ s^{-1}\ Mpc^{-1}}$.}
\tabcolsep 35.6pt%
\begin{tabular}{c c c c}
\toprule
Data&data0&data1&data2\\
\hline
$\Omega_bh^2$&$0.02242\pm0.00014$&$0.02243\pm0.00012$&$0.02242\pm0.00011$\\
$\Omega_ch^2$&$0.1174\pm0.0010$&$0.11740\pm0.00081$&$0.11743^{+0.00068}_{-0.00058}$\\
$100\theta_{\rm{MC}}$&$1.04103\pm0.00029$&$1.04103\pm0.00028$&$1.04102\pm0.00026$\\
$\tau$&$0.094\pm0.017$&$0.094\pm0.016$&$0.094\pm0.016$\\
${\rm{ln}}(10^{10}A_s)$&$3.118\pm0.032$&$3.118\pm0.032$&$3.118\pm0.032$\\
$n_s$&$0.9707\pm0.0041$&$0.9708\pm0.0037$&$0.9707\pm0.0034$\\
\hline
$\Omega_m$&$0.3075\pm0.0061$&$0.3072\pm0.0041$&$0.3073\pm0.0025$\\
$H_0$&$67.78\pm0.46$&$67.80\pm0.30$&$67.80\pm0.16$\\
\hline
$\sum m_\nu$&$<0.182$&$<0.176$&$<0.170$\\
\bottomrule
\end{tabular}
\centering
\end{table*}

\begin{figure*}[!t]
\centering
\includegraphics[scale=0.50]{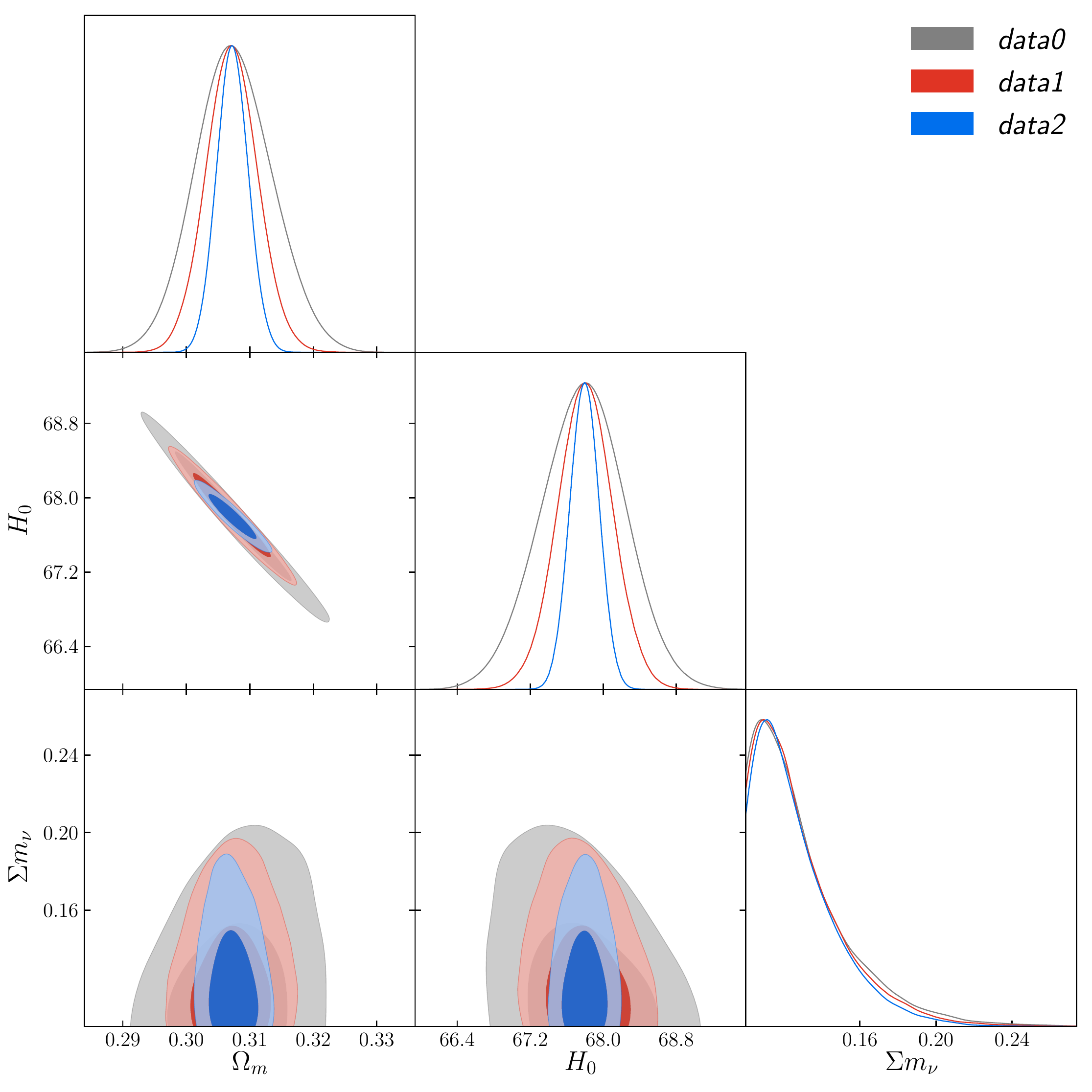}
 \caption{\label{fig2} The constraint results for the IH case of the neutrino mass ordering. Marginalized one- and two-dimensional posterior distributions of $\sum m_\nu$, $\Omega_{\rm m}$, and $H_0$ using data0, data1, and data2. Here, data0 denotes CMB+BAO+SN+$H_0$, data1 denotes CMB+BAO+SN+$H_0$+SKA1, and data2 denotes CMB+BAO+SN+$H_0$+SKA2. Note that $\sum m_\nu$ is in units of ${\rm eV}$ and $H_0$ is in units of ${\rm km\ s^{-1}\ Mpc^{-1}}$.}
\end{figure*}

\begin{table*}[t]
\footnotesize
\caption{\label{tab3}The constraint results for the DH case of the neutrino mass ordering. Here, data0 denotes CMB+BAO+SN+$H_0$, data1 denotes CMB+BAO+SN+$H_0$+SKA1, and data2 denotes CMB+BAO+SN+$H_0$+SKA2. Note that $\sum m_\nu$ is in units of ${\rm eV}$ and $H_0$ is in units of ${\rm km\ s^{-1}\ Mpc^{-1}}$.}
\tabcolsep 35.6pt%
\begin{tabular}{c c c c}
\toprule
Data&data0&data1&data2\\
\hline
$\Omega_bh^2$&$0.02238\pm0.00014$&$0.02238\pm0.00012$&$0.02238\pm0.00011$\\
$\Omega_ch^2$&$0.1182\pm0.0010$&$0.11817^{+0.00087}_{-0.00079}$&$0.11821^{+0.00076}_{-0.00059}$\\
$100\theta_{\rm{MC}}$&$1.04099\pm0.00030$&$1.04099\pm0.00028$&$1.04099\pm0.00027$\\
$\tau$&$0.086\pm0.017$&$0.086\pm0.016$&$0.086\pm0.016$\\
${\rm{ln}}(10^{10}A_s)$&$3.104\pm0.033$&$3.104\pm0.032$&$3.104\pm0.032$\\
$n_s$&$0.9688\pm0.0040$&$0.9687\pm0.0037$&$0.9687\pm0.0035$\\
\hline
$\Omega_m$&$0.3029\pm0.0060$&$0.3025\pm0.0040$&$0.3025\pm0.0024$\\
$H_0$&$68.24\pm0.47$&$68.27\pm0.30$&$68.27\pm0.16$\\
\hline
$\sum m_\nu$&$<0.109$&$<0.098$&$<0.091$\\
\bottomrule
\end{tabular}
\centering
\end{table*}

\begin{figure*}[!t]
\centering
\includegraphics[scale=0.50]{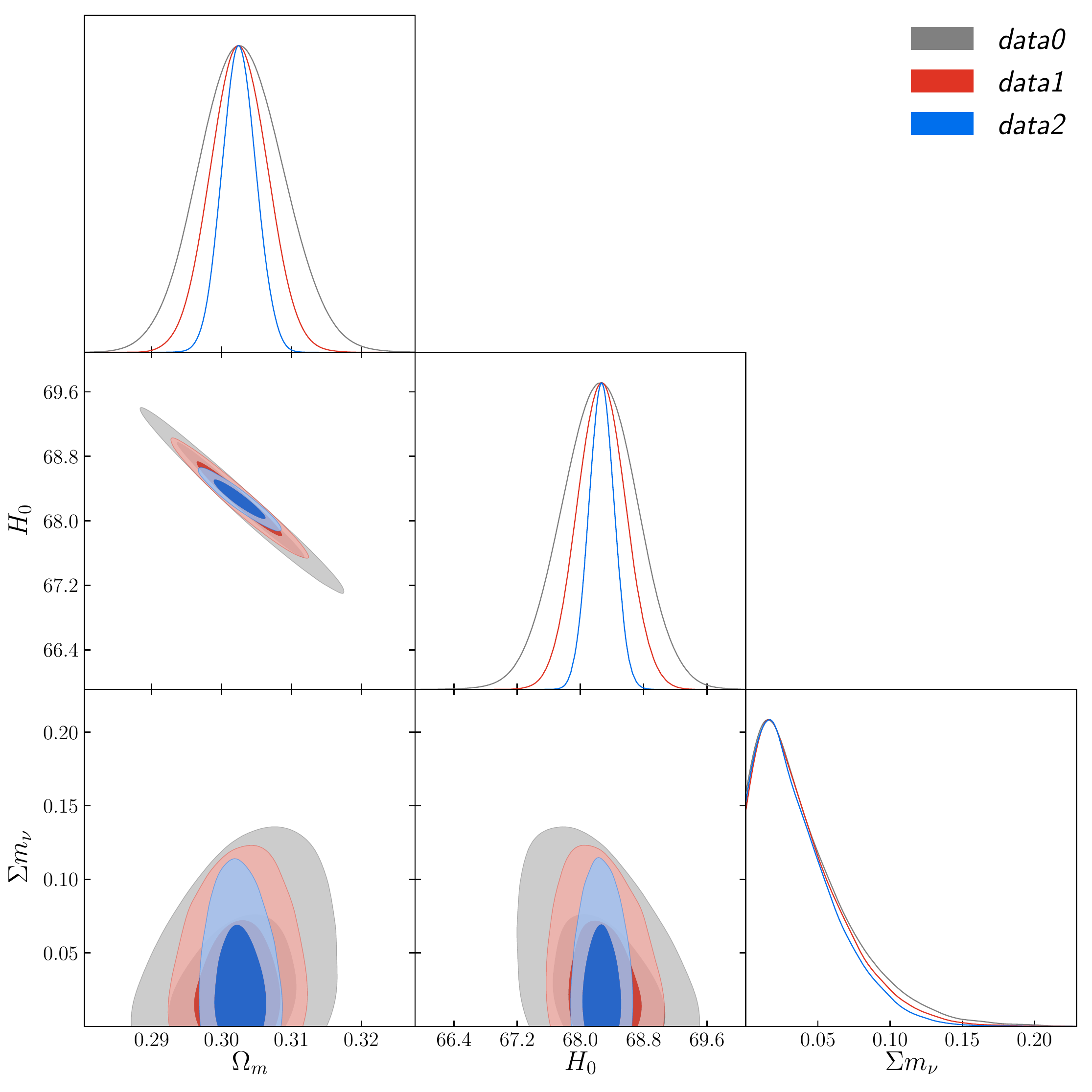}                                                                                                                                               \caption{\label{fig3} The constraint results for the DH case of the neutrino mass ordering. Marginalized one- and two-dimensional posterior distributions of $\sum m_\nu$, $\Omega_{\rm m}$, and $H_0$ using data0, data1, and data2. Here, data0 denotes CMB+BAO+SN+$H_0$, data1 denotes CMB+BAO+SN+$H_0$+SKA1, and data2 denotes CMB+BAO+SN+$H_0$+SKA2. Note that $\sum m_\nu$ is in units of ${\rm eV}$ and $H_0$ is in units of ${\rm km\ s^{-1}\ Mpc^{-1}}$.}
\end{figure*}

\section{Results and discussion}\label{sec:3}
In this section, we report the constraint results and make some relevant discussions. Our main results are shown in Tables~\ref{tab1}--\ref{tab3} and Figures~\ref{fig1}--\ref{fig3}. The results of the NH case are shown in Table~\ref{tab1} and Figure~\ref{fig1}, the results of the IH case are shown in Table~\ref{tab2} and Figure~\ref{fig2}, and the results of the DH case are shown in Table~\ref{tab3} and Figure~\ref{fig3}. Note also that here we use data0 to denote the data combination CMB+BAO+SN+$H_0$, use data1 to denote the data combination CMB+BAO+SN+$H_0$+SKA1, and use data2 to denote the data combination CMB+BAO+SN+$H_0$+SKA2.

From these results, we can clearly see that the SKA observations can significantly improve the constraints on all the parameters, in particular for the parameters $\Omega_{\rm m}$ and $H_0$. Compared to the SKA1 data, the SKA2 data have a much more powerful constraint capability. In the following, we will first report the constraint results of $\Omega_{\rm m}$ and $H_0$, and then discuss the constraint results of $\sum m_\nu$.

For the NH case, using data0 we have $\Omega_{\rm m}=0.3054\pm0.0061$ and $H_0=67.99\pm0.47\ {\rm km\ s^{-1}\ Mpc^{-1}}$; using data1 we have $\Omega_{\rm m}=0.3052\pm0.0041$ and $H_0=68.00\pm0.30\ {\rm km\ s^{-1}\ Mpc^{-1}}$; and using data2 we have $\Omega_{\rm m}=0.3052\pm0.0025$ and $H_0=68.00\pm0.16\ {\rm km\ s^{-1}\ Mpc^{-1}}$.

For the IH case, using data0 we have $\Omega_{\rm m}=0.3075\pm0.0061$ and $H_0=67.78\pm0.46\ {\rm km\ s^{-1}\ Mpc^{-1}}$;
using data1 we have $\Omega_{\rm m}=0.3072\pm0.0041$ and $H_0=67.80\pm0.30\ {\rm km\ s^{-1}\ Mpc^{-1}}$;
and using data2 we have $\Omega_{\rm m}=0.3073\pm0.0025$ and $H_0=67.80\pm0.16\ {\rm km\ s^{-1}\ Mpc^{-1}}$.

For the DH case, using data0 we have $\Omega_{\rm m}=0.3029\pm0.0060$ and $H_0=68.24\pm0.47\ {\rm km\ s^{-1}\ Mpc^{-1}}$;
using data1 we have $\Omega_{\rm m}=0.3025\pm0.0040$ and $H_0=68.27\pm0.30\ {\rm km\ s^{-1}\ Mpc^{-1}}$;
and using data2 we have $\Omega_{\rm m}=0.3025\pm0.0024$ and $H_0=68.27\pm0.16\ {\rm km\ s^{-1}\ Mpc^{-1}}$.

We find that using the same data combination, the constraints on $\Omega_{\rm m}$ and $H_0$ are similar for all the mass ordering cases. For the parameter $\Omega_{\rm m}$, the current observation CMB+BAO+SN+$H_0$ gives its error around 0.0060; when the SKA1 observation is added, the error becomes around 0.0040, and the constraint is improved by 33.3\%; and when the SKA2 observation is added, the error becomes around 0.0025, and the constraint is improved by 58.3\%.
For the parameter $H_0$, the current observation CMB+BAO+SN+$H_0$ gives its error around 0.47 km s$^{-1}$ Mpc$^{-1}$; when the SKA1 observation is added, the error becomes around 0.30 km s$^{-1}$ Mpc$^{-1}$, and the constraint is improved by 36.2\%; and when the SKA2 observation is added, the error becomes around 0.16 km s$^{-1}$ Mpc$^{-1}$, and the constraint is improved by 66.0\%.

Now we discuss the constraints on the neutrino mass. For the NH case, we have $\sum m_\nu<0.148\ {\rm eV}$, $\sum m_\nu<0.142\ {\rm eV}$, and $\sum m_\nu<0.137\ {\rm eV}$ by using data0, data1, and data2, respectively. For the IH case, we have $\sum m_\nu<0.182\ {\rm eV}$, $\sum m_\nu<0.176\ {\rm eV}$, and $\sum m_\nu<0.170\ {\rm eV}$ by using data0, data1, and data2, respectively.
For the DH case, we have $\sum m_\nu<0.109\ {\rm eV}$, $\sum m_\nu<0.098\ {\rm eV}$, and $\sum m_\nu<0.091\ {\rm eV}$ by using data0, data1, and data2, respectively. Here, as usual, the upper limit values of the neutrino mass refer to the 95.4\% ($2\sigma$) confidence level. Note also that the lower bounds of the total neutrino mass have been set for the NH case and the IH case, which are 0.06 and 0.10 eV, respectively, as shown in Figures~\ref{fig1} and \ref{fig2}. For the DH case, there is no a lower bound of the total neutrino mass; see also Figure~\ref{fig3}.

We find that, for the constraints on the neutrino mass $\sum m_\nu$, when the SKA1 observation is added, the upper limit values are reduced by $4.1\%$, $3.3\%$, and $9.8\%$ for the NH, IH, and DH cases, respectively; when the SKA2 observation is added, the upper limit values are reduced by $7.4\%$, $6.5\%$, and $16.4\%$ for the NH, IH, and DH cases, respectively. Therefore, it is found by this investigation that the SKA observations can also improve the constraints on the total neutrino mass to some extent.

The issue of how SKA 21 cm IM observation can contribute to the cosmological constraints on the sum of the neutrino masses has been previously discussed in Refs.~\cite{Oyama:2015gma,Olivari:2017bfv,Sprenger:2018tdb}. In Ref.~\cite{Oyama:2015gma}, the authors use the simulated future observations of 21 cm line radiation coming from the epoch of reionization (EoR) to discuss the constraints on the neutrino mass by calculating the Fisher matrix, and they find that by combining the precise CMB polarization observation from Simons Array with the 21 cm line observation from SKA1 and the BAO observation from DESI, the effects of non-zero neutrino mass on the growth of density fluctuation can be measured, if the total neutrino mass is larger than 0.1eV. In Refs.~\cite{Olivari:2017bfv,Sprenger:2018tdb}, the authors use the full theoretical power spectrum instead of just BAO information from the 21 cm IM observation to derive constraints, examine synergies with other surveys and discuss the effects of additional free parameters. Comparing with these previous studies can validate our new results.  The differences between our work and these previous studies are mainly embodied in the following points. (i) We use the SKA mid- and high-frequency observations, not the low-frequency of EoR observation. (ii) We use the HI BAO measurements, but not the full HI power spectrum, to make cosmological parameter constraints. (iii) In the cosmological fit, we use the MCMC method, instead of the Fisher matrix method. (iv) In the cosmological analysis, we do not assume a detection of the neutrino mass, but only consider upper limits on the neutrino mass. (v) We investigate how the 21 cm observations from SKA can improve the constraints on cosmological parameters (in particular, the neutrino mass) on the basis of the constraints from the current CMB+BAO+SN+$H_0$ data. Note here that using the current mainstream cosmological observations (CMB+BAO+SN+$H_0$), the cosmological parameters have been tightly constrained. (vi) We make discussion for
the NH, IH, and DH cases, respectively. Here, it should be emphasized that we do not use the 21 cm observation to distinguish the neutrino mass ordering, but only consider the constraints on the total neutrino mass in the three cases of neutrino mass ordering.

Although the total neutrino mass has still not been determined, the upper limits on it have been rather stringent using the current CMB+BAO+SN+$H_0$ data. In this work, we show that, using the future 21 cm observations from the SKA, the upper limits on the total neutrino mass can be further decreased by about 4\% (NH), 3\% (IH), and 10\% (DH) for SKA1, and by about 7\% (NH), 7\% (IH), and 16\% (DH) for SKA2. Here, we assume that the total neutrino mass has still not been determined by the observations. Of course, in the future, with the helps from the 21 cm observation and other survey observations, the total neutrino mass is rather likely to be precisely determined.
\section{Conclusion}\label{sec:4}
In this work, we investigate how the SKA HI 21-cm IM and galaxy sky surveys can be used to improve the constraints on the total neutrino mass. We use the simulated data of the BAO measurements from the HI sky survey based on SKA1 (IM) and SKA2 (galaxy) to do the analysis. We wish to see, compared to the current observations, what role the SKA observations can play in the cosmological parameter estimation in a cosmological model involving massive neutrinos. For the current observations, we use the {\it Planck} 2015 CMB data, the BAO data, the SN data (Pantheon compilation), and the $H_0$ measurement. We use three data combinations, CMB+BAO+SN+$H_0$, CMB+BAO+SN+$H_0$+SKA1, and CMB+BAO+SN+$H_0$+SKA2, to perform constraints, and to further make a comparison. We consider three mass ordering cases for massive neutrinos, i.e., the NH, IH, and DH cases.

We find that the SKA observation can significantly improve the constraints on $\Omega_{\rm m}$ and $H_0$. Compared to the current observation, the SKA1 data can improve the constraints on $\Omega_{\rm m}$ by about 33\%, and on $H_0$ by about 36\%; the SKA2 data can improve the constraints on $\Omega_{\rm m}$ by about 58\%, and on $H_0$ by about 66\%.

We find that the SKA observation can also slightly improve the constraints on $\sum m_\nu$. Compared to the current observation, the SKA1 data can improve the constraints on $\sum m_\nu$ by about 4\%, 3\%, and 10\%, for the NH, IH, and DH cases, respectively; the SKA2 data can improve the constraints on $\sum m_\nu$ by about 7\%, 7\%, and 16\%, for the NH, IH, and DH cases, respectively.

It is expected that in the future the SKA observation, combined with the future highly accurate optical survey projects, such as LSST, Euclid, and WFIRST, as well as the gravitational-wave standard siren observations from ground-based and space-based detectors, would greatly promote the development of cosmology.

\Acknowledgements{This work was supported by the National Natural Science Foundation of China (Grants Nos.~11875102, 11835009, 11975072, 11690021, and 11522540), the Liaoning Revitalization Talents Program (Grant No.~XLYC1905011), the Fundamental Research Funds for the Central Universities (Grant No.~N2005030), and the National Program for Support of Top-Notch Young Professionals.}





\end{multicols}
\end{document}